\documentstyle[psfig]{mn}                

\title[W~Crv]
{W~Crv: The shortest-period Algol with non-degenerate components?
\thanks{Based on observations obtained at the David Dunlap
Observatory, University of Toronto.}}

\author[Rucinski and Lu]
{ Slavek M. Rucinski and Wenxian Lu \\
David Dunlap Observatory, University of Toronto,
   Box 360, Richmond Hill, Ontario L4C~4Y6, Canada\\
{\rm e-mail:} rucinski@astro.utoronto.ca, lu@astro.utoronto.ca}

\date{Accepted 
      Received 
      in original form 1999 July }  
 
\pagerange{\pageref{firstpage}--\pageref{lastpage}}
\pubyear{1999}
 
\begin{document}
 
\maketitle

\label{firstpage}

\begin{abstract}
Radial velocity data for both components of W~Crv are presented.
In spite of providing full radial-velocity information,
the new data are not sufficient to 
establish the configuration of this important system
because of large seasonal light-curve variations which 
prevent a combined light-curve/radial-velocity solution. It is 
noted that
the primary minimum is free of the photometric variations, a property
which may help explain their elusive source. Photometrically, the 
system appears to be a contact binary with poor or absent energy
exchange, but such an explanation -- in view of the presence of the
mass-transfer effects -- is no more plausible than any one
of the semi-detached configurations with either the 
more-massive or less-massive components filling their Roche 
lobes. Lengthening of the orbital period and the size
of the less-massive component above its main-sequence value
suggest that the system is the shortest-period (0.388 days)
known Algol with non-degenerate components.
\end{abstract}


\section{Introduction}

Contact binary stars are common: According to the only currently
available unbiased statistics -- 
a by-product of the OGLE microlensing
project -- as discussed in Rucinski \shortcite{ruc97a} 
and Rucinski \shortcite{ruc98b}, 
the spatial frequency of contact binaries among the main-sequence,
galactic-disk
stars of spectral types F to K (intrinsic colors $0.4 < V-I_C < 1.4$)
is about 1/100 to 1/80 (counting contact binaries as single
objects, not as two stars). Most of them have orbital periods within
$0.25 < P < 0.7$ days, and they are very rare for
$P > 1.3 - 1.5$ days \cite{ruc98a}. These properties, as 
well as the spatial distribution extending all the way to the 
galactic bulge, with moderately large $z$ distances from the 
galactic plane, and the kinematic properties \cite{GB88} suggest an Old
Disk population of Turn-Off-Point binaries, i.e.\ a population
characterized by conditions conducive to rapid synchronization and
formation of contact systems from close, but detached, binaries. 
The contact binaries are less common in open clusters which are
younger than the galactic disk \cite{ruc98b}, a property indicating
that they form over time of a few Gyrs. It is
obviously of great interest to identify binaries which are related to,
or precede the contact system stage, as the relative numbers would
give us information on durations of the pre- and in-contact stages.

Lucy \shortcite{luc76} and Lucy \& Wilson
\shortcite{LW79} were the first to point out 
the observational importance of contact
systems with unequally deep eclipses as possible
exemplification of binaries which are to become contact systems or
are in the ``broken-contact'' phase of the theoretically predicted
Thermal Relaxation Oscillation (TRO) evolution of contact
binary stars, as discussed by Lucy \shortcite{luc76},
Flannery \shortcite{fla76} and 
Robertson \& Eggleton \shortcite{RE77}. Lucy \& Wilson called 
such contact systems the B-type -- as contrasted to the 
previously recognized W-type and A-type contact systems -- 
because of the light curves 
resembling those of the $\beta$~Lyrae-type binaries. 
While the A-type are the
closest to the theoretical model of contact binaries with perfect
energy exchange and temperature equalization, the W-type show 
relatively small (but still unexplained)
deviations in the sense that {\it less-massive components\/}
have slightly higher surface brightnesses (or temperatures). 
Systems of the B-type introduced by Lucy \& Wilson
show large deviations from the contact model in that 
{\it more massive components\/} are hotter than predicted by 
the contact model. Thus, the energy transfer is inhibited 
or absent and the components of the B-type systems behave more like 
independent (or thermally de-coupled) ones. While 
light-curve-synthesis solutions suggest good geometrical 
contact, it has been suggested that these may be semi-detached
binaries with hotter, presumably more-massive
components filling their Roche lobes (we will call
these SH following Eggleton \shortcite{egg96}). 

The same OGLE statistics that gave indications of the very high spatial
frequency of contact binaries suggests that
short-period binaries which simultaneously are in contact and
show unequally-deep eclipses are relatively rare in space:
Among 98 contact systems in the volume
limited sample, only 2 have unequally deep minima indicating
components of different effective temperatures \cite{ruc97b}. 
Both of these
systems (called there ``poor-thermal-contact'' or ``PTC'' systems, 
but which could be as well called B-type contact systems)
have periods longer than 0.37 day and both show the first 
maximum (after the deeper eclipse) relatively higher of 
the two maxima. This type of
asymmetry is dominant in the spatially much larger (magnitude limited)
sample of systems available in the OGLE survey. As already pointed
by Lucy \& Wilson \shortcite{LW79}, this sense of asymmetry can be 
explained most easily as a manifestation of mass-transfer 
from the more-massive to the less-massive component.
We add here that this can happen also in a {\it non-contact\/} 
SH system, 
with the continuum light emission from the interaction volume
between stars contributing to the strong curvature of the light-curve
maxima and mimicking the photometric effects 
of the tidally-elongated (contact) structure. 
Exactly this type of asymmetry is observed in a system which
is absolutely crucial in the present context, V361~Lyr;  
it has been studied by Ka\l u\.zny \shortcite{kal90} 
and Ka\l u\.zny \shortcite{kal91}, and later convincingly shown by 
Hilditch et al.\ \shortcite{hil97} to be 
a semi-detached binary with matter flowing from the more massive to the
less-massive component. The light curve asymmetry in the case of
V361~Lyr is particularly large and stable. A similar asymmetry
and somewhat similar mass-transfer effects (albeit involving 
much more massive
components) are observed in the early-type system SV~Cen \cite{ruc92}
where we have a direct evidence of a tremendous mass-transfer 
in a very large period change.

The subject of this paper, the close binary  
W~Crv (GSC 05525--00352, BD$-12$~3565) is a relatively
bright ($V=11.1$, $B-V=0.66$) system with the
orbital period of 0.388 day. For a long time, this was the short-period
record holder among systems which appear to be
in good geometrical contact, yet which show strongly 
unequally-deep eclipses indicating poor thermal contact. 
It was as one of the systems exemplifying the 
definition of contact systems of the B-type
by Lucy \& Wilson \shortcite{LW79}, although most often its
type of variability has been characterized as EB or $\beta$ Lyrae-type.  
A system photometrically
similar to W~Crv with the period of 0.37 days, \#3.012, 
has been identified in the OGLE sample \cite{ruc97b}, but it is
too faint for spectroscopic studies. 

Our radial velocity data which we describe in this paper
are the first spectroscopic results for W~Crv. 
Thus, it would be natural to combine them with the previous
photometric studies. However, we will claim below 
that W~Crv is more complex than the current 
light-curve synthesis codes can handle. The previous analyses of the
system, without any spectroscopic constraints on
the mass-ratio ($q$), encountered severe
difficulties. A recent extensive study of several light curves of
W~Crv by Odell \shortcite{ode96}, solely based on
photometric data found that the
mass-ratio was practically indeterminable ($0.5 < q < 2$), admitting
solutions ranging between the Algol systems (SC, for semi-detached 
with the cool, lower-mass component filling its Roche lobe)
on one hand and all possible configurations which are
conventionally used to explain
the B-type light curves (SH, i.e.\ the broken-contact or pre-contact 
semi-detached systems as well as poor-thermal-contact systems)
on the other hand.
A value of $q=0.9$ and the more massive component being eclipsed
at primary minimum were assumed by Odell mostly by plausibility arguments.

For a comprehensive summary of the theoretical issues related 
to pre- and in-contact evolution, the reader is suggested to
refer to the review of Eggleton \shortcite{egg96}; 
observational data for B-type systems similar to W~Crv were collected
and discussed in a five-part series by Ka\l u\.zny, concluded
with Ka\l u\.zny \shortcite{kal86}, and in studies by
Hilditch \& King \shortcite{hil86}, Hilditch et al.\ \shortcite{hil88} 
and Hilditch \shortcite{hil89}.

\section{Radial velocity observations}

\begin{figure}   
\centerline{\psfig{figure=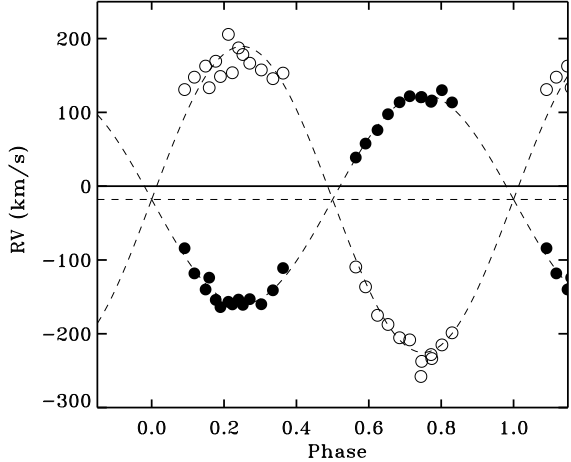,height=6cm}}
\vspace{0.25cm}
\caption{\label{fig1}
The radial velocity observations of W~Crv versus the
orbital phase. The hotter, more massive component eclipsed in the primary
minimum is marked by filled circles. The data are listed in
Table~\ref{tab1} and the sine-curve
fits (broken lines) correspond to elements given in Table~\ref{tab2}.}
\end{figure}

The radial velocity observations of W~Crv were 
obtained in February -- April 1997
at David Dunlap Observatory, University of Toronto using the 1.88 metre
telescope and a Cassegrain spectrograph. 
The spectral region of 210 \AA\ centered
on 5185 \AA\ was observed at the spectral
scale of 0.2 \AA/pixel or 12 km~s$^{-1}$/pixel. The 
entrance slit of the spectrograph of 1.8 arcsec on the sky 
was projected into about 3.5 pixels or 42 km~s$^{-1}$. The exposure
times were typically 10 to 15 minutes.
The radial velocity data are listed in Table~\ref{tab1}
and are shown graphically in Figure~\ref{fig1}. The component velocities
have been determined by fitting gaussian curves to peaks in 
the broadening function obtained
through a de-convolution process, as described in Lu \& Rucinski 
\shortcite{LR99}. The mean standard deviations from the sine-curve
variations are 7.7 km~s$^{-1}$ for the primary (more-massive,
subscript 1) component and 17.2 km~s$^{-1}$ for the secondary 
(less-massive, subscript 2) component. These deviations
give the upper limits to the measurement uncertainties because they
contain the deviations of the component velocities 
from the simplified model of circular orbits without any proximity
effects (i.e.\ without allowance for non-coinciding photometric
and dynamic centres of the components).

\begin{table}    
\caption{\label{tab1} Radial velocity observations of W~Crv}
\begin{tabular}{@{}rrrrrr}
JD(hel)  & Phase& V$_{pri}$ & O--C   & V$_{sec}$ & O--C    \\
2450000+&       &   km~s$^{-1}$    & km~s$^{-1}$   &  km~s$^{-1}$     &         \\
489.811 & 0.564 & 38.8      &	1.7	 & $-109.5$  & $-10.9$ \\
489.822 & 0.591 & 57.7	    & $-0.9$ & $-136.4$	 & $-6.3$  \\
489.835 & 0.624 & 76.0	    & $-5.9$ & $-175.0$	 & $-11.0$ \\
489.846 & 0.653 & 97.5	    & $-0.6$ & $-187.4$	 & 0.4  \\
489.858 & 0.685 &	113.7	    & 1.6	 & $-205.4$	 & 2.9  \\
489.869 & 0.713 &	121.9	    & 1.9	 & $-208.3$	 & 11.5 \\
489.881 & 0.744 &	120.4	    & $-3.4$ & $-257.9$	 & $-32.6$ \\
489.892 & 0.772 &	114.3	    & $-8.2$ & $-228.2$	 & $-4.7$  \\
489.903 & 0.802 &	130.1	    & $13.6$ & $-215.1$	 & $-0.5$  \\
489.914 & 0.830 &	113.5	    & $7.1$	 & $-198.7$	 & 1.2  \\
520.674 & 0.091 &	$-84.1$   & $10.3$ & 130.9	 & 37.2 \\
520.684 & 0.118 &	$-118.2$  & $-4.5$ & 147.7	 & 25.8 \\
520.697 & 0.149 &	$-140.0$  &	$-7.6$ & 162.7	 & 13.5 \\
520.707 & 0.177 &	$-154.1$  &	$-8.8$ & 169.3	 & 1.3 \\
520.721 & 0.212 &	$-156.7$  &	$-0.8$ & 205.7	 & 22.1 \\
520.732 & 0.240 &	$-153.7$  &	5.9	 & 187.4	 & $-1.6$ \\
520.744 & 0.271 &	$-153.1$  &	5.6	 & 166.6	 & $-21.0$ \\
520.756 & 0.303 &	$-160.0$  &	$-7.9$ & 157.5	 & $-20.5$ \\
520.769 & 0.335 &	$-141.1$  & $-0.8$ & 145.7	 & $-15.1$ \\
520.779 & 0.363 &	$-111.0$  &	14.8	 & 153.2	 & 13.6 \\
535.675 & 0.746 &	120.8	    & $-3.0$ & $-237.4$	 & $-12.1$ \\
535.686 & 0.774 &	116.1	    & $-6.2$ & $-233.6$	 & $-10.5$ \\
539.716 & 0.159 &	$-124.0$  & 13.3	 & 133.4	 & $-22.9$ \\
539.728 & 0.190 &	$-163.9$  &	$-14.0$& 148.6	 & $-26.2$ \\
539.741 & 0.223 &	$-160.2$  &	$-2.4$ & 153.7	 & $-32.6$ \\
539.752 & 0.252 &	$-160.8$  & $-0.9$ & 178.4	 & $-11.0$ \\ 
\end{tabular}
\end{table}

The individual observations as well as the
{\it observed minus calculated\/} $(O-C)$ deviations from the 
sine-curve fits to radial velocities of individual components 
are given in Table~\ref{tab2}. When finding the parameters of
the fits, we assumed only the value of the
period, following Odell \shortcite{ode96},
and determined the mean velocity $V_0$,
the two amplitudes $K_1$ and $K_2$ as well as the moment of the
primary minimum $T_0$. The remaining quantities in that table have
been derived from the amplitudes $K_i$. The errors of the
parameters have been determined by a bootstrap experiment
based on 10,000 solutions with randomly selected observations 
with repetitions.

\begin{table}    
\caption{\label{tab2} Circular orbit solution for W~Crv}
\begin{tabular}{@{}cccc}
Parameter & Units   & Value     & Comment \\
$T_0$     & JD(hel) & $2450489.9781 \pm 0.0015$ & \\
$P$       & days    & 0.388081  & assumed \\
$K_1$     & km~s$^{-1}$    & $140.8 \pm 2.0$ & \\
$K_2$     & km~s$^{-1}$    & $206.4 \pm 3.7$ & \\
$V_0$     & km~s$^{-1}$    & $-20.1 \pm 1.8$ & \\
$q$       &         & $0.682 \pm 0.016$ & derived\\
$(a_1+a_2) \sin i$  & $R_\odot$ & $2.66 \pm 0.04$ & derived \\
$M_1 \sin^3 i$ & $M_\odot$  & $1.00 \pm 0.06$ & derived \\
$M_2 \sin^3 i$ & $M_\odot$  & $0.68 \pm 0.05$ & derived \\  
\end{tabular}
\end{table}

Among the spectroscopic elements in Table~\ref{tab2}, 
the mass-ratio, $q = 0.682 \pm 0.016$,
is the most important datum for proper interpretation of the
light curves. Without external information
on the mass-ratio, strong inter-parametric correlations
in the light-curve analyses are known to frequently produce entirely
wrong solutions (except for cases of total eclipses). 

Before attempting a  combined solution,
we note that the spectroscopic data, as given in
Table~\ref{tab2}, describe the following system:
The more-massive component is eclipsed in the deeper
eclipse and hence is the hotter of the two. Judging by the
relative depths of the eclipses, and noting the small light
contribution of the secondary component (even if it fills its 
Roche lobe), we estimate -- on the basis of the systemic colour at light
maxima $(B-V)=0.66$ -- that the effective temperatures of the
components are approximately 5700K and 4900K. 
The mass of the primary component is 
$M_1\,\sin^3 i = 1.00\, M_\odot$, so that the primary is apparently 
a solar-type star, and the orbital inclination
cannot be far from $i=90^\circ$, although not exactly so as total
eclipses are not observed. Obviously,
the spectroscopic data cannot provide any constraint on the degree of
contact in the system, i.e.\
whether it is a contact system with poor 
thermal contact or a semi-detached configuration with one of 
the components filling the Roche lobe or perhaps even
a detached binary. There are no spectroscopic
indications of any mass-transfer either, 
although -- with the mutual proximity of
components -- one would not expect such obvious signatures of this
process as a stream or an accretion disk; besides, the spectral
region around 5185 \AA\ would not normally show them in any case.
We must seek for constraints on the 
system geometry in the light curve and its variations.

\begin{figure}   
\centerline{\psfig{figure=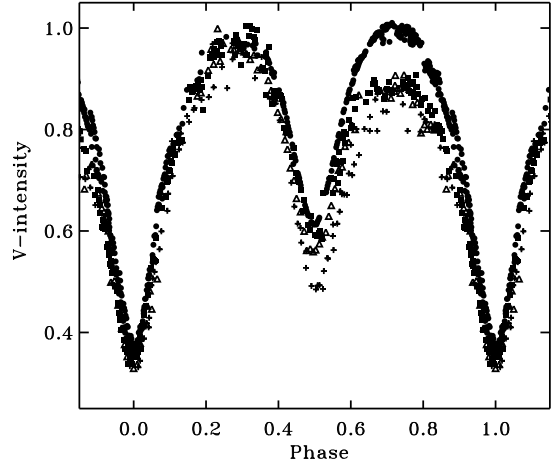,height=6cm}}
\vspace{0.25cm}
\caption{\label{fig2} Four seasonal V-filter light curves 
of W~Crv as discussed by Odell \shortcite{ode96} 
are shown here together, in intensity units, assuming the difference
of 0.37 mag between the comparison and the variable star. 
Note the good repetition of the light curves in primary minima and large
variations elsewhere. The codes are: 1966 -- crosses, 1981 -- filled
circles, 1988 -- filled squares, 1993 -- triangles.}
\end{figure}

\section{Attempts of a combined light curve and radial velocity solution}

Four light curves discussed by Odell \shortcite{ode96} 
are currently available: the first from 1966 
was obtained by Dycus \shortcite{dyc68}, the 
remaining three in 1981, 1988 and 1993 were by Odell.
The light curves were
obtained with the same comparison star permitting direct
comparison of the large curves. The large seasonal variations 
of the light curves were 
interpreted by Odell by star spots. We do not support the spot
hypothesis by pointing out a curious property: A comparison
of the seasonal light curves (Figure~\ref{fig2}) indicates 
that all changes take place at light maxima and during the
secondary eclipse when the cooler component is behind the hotter one,
but that primary eclipse is surprisingly similar in all four curves.
This constancy of the primary-eclipse shape
remains irrespectively whether one considers the intensity 
or magnitude (relative intensity) units.
We feel that we have here a strong indication that
mass-exchange and accretion processes are operating
between the stars. These processes would produce large
areas of hot plasma, most probably on the inner face of the
less-massive secondary component which is invisible during the primary
minima. One can of course contrive a scenario involving dark spots
appearing in certain areas, but never appearing
on the outer side of the less-massive component, but the dark-spot
hypothesis seems to be the most artificial of all possibilities. We note
that an argument of the diminished brightness being 
accompanied by a redder
colour is a weak one as such correlation is expected when plasma
temperature effects are involved, irrespectively whether the spots
are cool or hot.

With strong mass-transfer effects modifying its light curve, 
W~Crv is not a typical contact system. In this situation, 
a blind application of light-curve synthesis codes may have 
led us to entirely wrong sets of parameters.
For that reason, we did not attempt to obtain a light-curve solution
of the system and used the popular light-curve synthesis
program BinMak2 (as described by Bradstreet \shortcite{bra94} and Wilson
\shortcite{wil94}) to explore reasonable 
ranges of parameters in different geometrical configurations. 

\begin{figure}   
\centerline{\psfig{figure=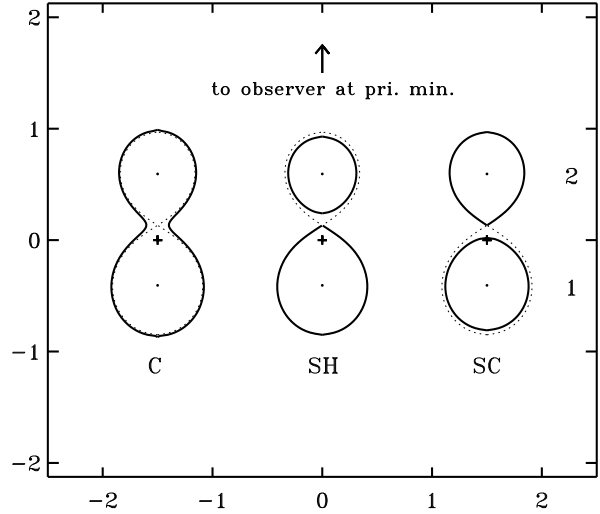,height=7cm}}
\vspace{0.25cm}
\caption{\label{fig3} The three configurations of W~Crv
considered in the text, with parameters as listed
in Table~\ref{tab3}, are shown here as sections in the orbital
plane. The Roche critical equipotentials (dotted
lines) and the position of the mass center 
(cross) are shown to scale. Note how little
space separates the components; this leads to our hypothesis that strong
mass-transfer phenomena between the components are the source of
additional light which produces the seasonal variations of the light
curve.
}
\end{figure}

\begin{figure}      
\centerline{\psfig{figure=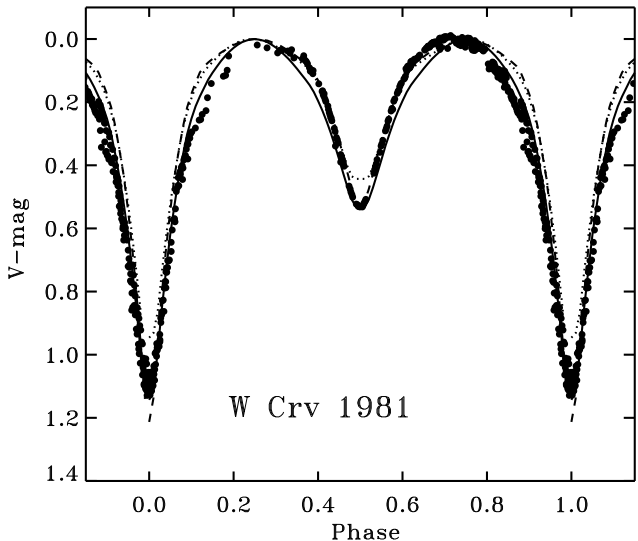,height=7cm}}
\vspace{0.25cm}
\caption{\label{fig4} The 1981 light curve is shown
here with the three fits: the 
contact model (C) with a mild degree-of-contact
($f=0.15$, continuous line) and two semi-detached 
configurations discussed in the text (SH, dotted line and
SC, broken line). 
}
\end{figure}

\begin{figure*}   
\centerline{\psfig{figure=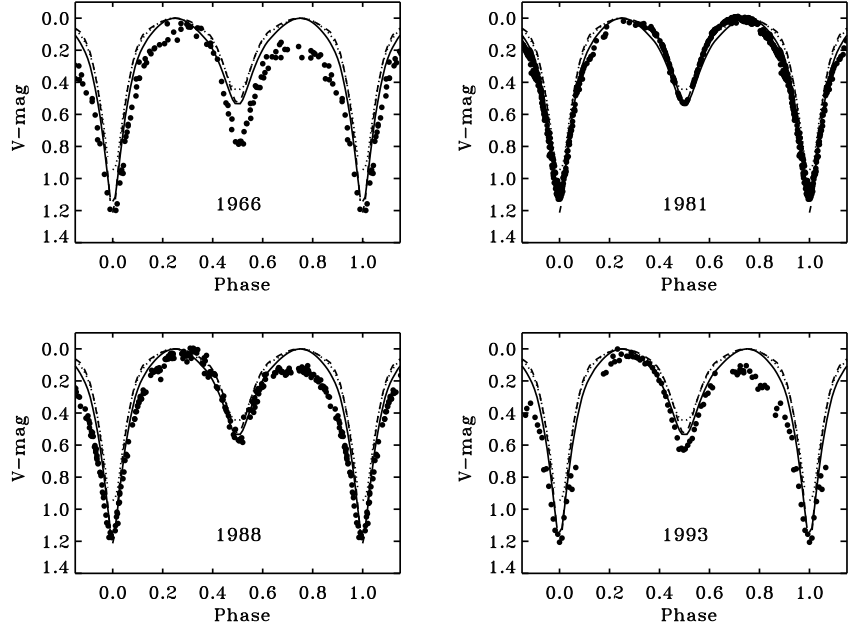,height=8cm,angle=90}}
\vspace{0.25cm}
\caption{\label{fig5} The four V-filter light curves 
of W~Crv (in magnitudes) are shown here together with three 
different fits for a contact model (C) with a mild degree-of-contact
($f=0.15$, continuous line), for two semi-detached 
configurations discussed in the text (SH, dotted line and
SC, broken line). The fits have been based on 
the 1981 light curve (see Figure~4).
Note the small differences between the theoretical
curves when compared with the large seasonal variations in the
observed light curves.
}
\end{figure*}

Attempts of conventional light-curve 
synthesis solutions of W~Crv encounter several problems.
First of all, the large amplitudes {\it at both minima\/} 
totally exclude a detached configuration.
At least one of the components or possibly both contribute to
the strong ellipticity of the light curve, which would not be
surprising in view of the short orbital period and little space
for expansion of components in the system. The system 
must be a contact one or must be described by
one of the two possible semi-detached configurations.
Arguably, durations of sub-contact phases of evolution are 
very short and the system should quickly reach a semi-detached stage. 
Let us call
the three possibilities ``C'' for contact, ``SH'' for the one
with the more massive component filling the Roche lobe and ``SC''
for an Algol configuration with the less massive component filling
its lobe. The shapes of the orbital cross-sections 
of the components for these three
possibilities are shown in Figure~\ref{fig3}. We will discuss 
them in turn, in reference to Figures~\ref{fig4} and \ref{fig5}
which show the most symmetric 1981 light curve and then
the four seasonal light curves. The 1981 curve was selected
for its relatively symmetric shape, good phase coverage 
and absence of what was initially thought to be signatures of dark spots. 

The parameters of the best-fitting synthesis models for the V-filter
1981 light curve are given in Table~\ref{tab3}. The values of
equipotentials $\Omega_i$ are defined as in the Wilson-Devinney 
program \cite{WD71}
and $r_i$ are the volume radii in units of the orbital centre
separation. The following assumptions on the properties of the
components of W~Crv were made while generating the
synthetic light curves: The limb darkening coefficients 
$u_1 = 0.65$ and $u_2 = 0.75$, the gravity exponents
$g_1 = g_2 = 0.32$ and the bolometric albedo $A_1 = A_2 = 0.5$.
The inner and outer equipotentials for $q = 0.682$ were
$\Omega_{in} = 3.215$ and $\Omega_{out} = 2.821$.
The radii given in Table~\ref{tab3} are the volume radii.

\begin{table}    
\caption{\label{tab3} Three light-curve synthesis solutions of W~Crv}
\begin{tabular}{@{}lccc}
Parameter  & C       & SH        & SC \\
$\Omega_1$ & 3.156 & 3.215       & 3.4 \\
$\Omega_2$ & 3.156 & 3.4         & 3.215 \\
$i$ (deg)  &  88     & 90        & 90  \\
$r_1$      & 0.424   & 0.412     & 0.380 \\
$r_2$      & 0.357   & 0.313     & 0.345 \\
$R_1/R_\odot \sin i$ & 1.13 & 1.10 & 1.01 \\
$R_2/R_\odot \sin i$ & 0.95 & 0.83 & 0.92 \\
Comment    & $f=0.15$ & primary  & secondary \\
           &          & fills R.\ lobe & fills R.\ lobe \\
\end{tabular}
\end{table}

{\bf Contact configuration (C):}
Conventional contact solutions make it
abundantly clear that the strong curvature of 
light maxima and large amplitude of light
variations require two properties: a large orbital inclination and a 
moderately strong contact, at least $f \simeq 0.15 - 0.25$. 
However, the inclination cannot
be exactly 90 degrees as then we would see a total eclipse
in the secondary minimum. The contact-model fit is far from
perfect because of the large seasonal changes, but also indicates a need 
of a ``super-reflection'' effect, with
increased albedo not only above the currently most popular value of 0.5
for convective envelopes \cite{ruc69},
but even above its physically allowed upper
limit of unity. This is clearly visible in Figures~\ref{fig4} and 
\ref{fig5} in the branches of the secondary minimum.
Cases of the abnormal reflection were already discussed by 
Lucy \& Wilson \shortcite{LW79} -- including the case of W~Crv --
and by Ka\l u\.zny \shortcite{kal86}, as indicating some
abnormal brightness distribution between the stars
(most probably, on the inner side of the 
secondary component) which could be linked to a mass-exchange phenomenon.
Obvious presence of such effects would make the standard, light-curve
synthesis model -- which
hides all energy and mass transfers deep inside the common 
contact envelope -- entirely invalid. 

{\bf Semi-detached configuration (SH):} This is the preferred
configuration for B-type systems, either in terms of
a system before forming contact or in the broken-contact phase
of the TRO oscillations. Photometrically, the model does not provide
enough of the light-curve amplitude and curvature at maxima,
even with $i=90^\circ$. The dotted line in Figures~\ref{fig4} and
\ref{fig5} shows this deficiency. However, in this 
configuration, it would be natural to expect
departures from the simple geometric model due to the mass exchange
phenomena. The increased reflection effect
could be then explained through an area
on the secondary component which is directly struck by the in-falling
matter from the primary component, while the strong curvature of maxima
could be explained by a light contribution from the accretion region
which is visible only at the quadratures, as is most likely the case
for SV~Cen \cite{ruc92}. Although such a configuration cannot
be modeled with the existing light-curve synthesis codes,
it offers a prediction of the shortening of the orbital period;
in Section~\ref{period} we present
indications that the period is in fact getting longer. 
It is also consistent with the 
light curve variations almost entirely limited to the light
maxima, with very small seasonal differences between portions
at light minima. If the mass-transfer phenomena between the stars
increase the light-curve amplitude, then the inclination 
could take basically any value. For $i < 90$ degrees, the inner side
of the secondary component would be partly visible at secondary
minima explaining large light-curve variations at these phases.

{\bf Semi-detached Algol configuration (SC):} Of the three 
geometrical models considered here,
this one best fits the 1981 light curve in all parts except 
in the upper branches of the
primary minimum which are wider than predicted. The large
amplitudes of the light variations find a better explanation
in this model than in the SH case. Also, most of the 
reflection effect
can be explained with the conventional value of the albedo
by the relatively larger area of the illuminated secondary component. 
The mass-transfer in this model should lead to
a period lengthening, as in other Algols. This is what we apparently
see in the times of minima of W~Crv
(see Section~\ref{period}). If the light-curve maxima contain
a light contribution of mass-transfer and/or accretion 
effects, then the second maximum
(after the secondary minimum) would be expected -- on the average -- 
to be more perturbed by the Coriolis-force deflected stream, 
and this seems to be the case for W~Crv
(see Figure~\ref{fig2}). Within the SC hypothesis, only
one of the two components, the secondary, would be abnormal (oversize
relative the main-sequence relation, see Tables~\ref{tab2}
and \ref{tab3}), whereas
the C and SH models predict mass-radius inconsistencies
for both components. Thus, we feel
that all the current data suggest that the short-period Algol
configuration is the correct explanation for W~Crv. The major problem,
however, is with the theoretical explanation for such a configuration:
There is simply no place for Algols with periods as short as
0.388 days within the present theories. We return to this problem in
Section~\ref{disc}.

\section{Period changes}
\label{period}

\begin{figure}   
\centerline{\psfig{figure=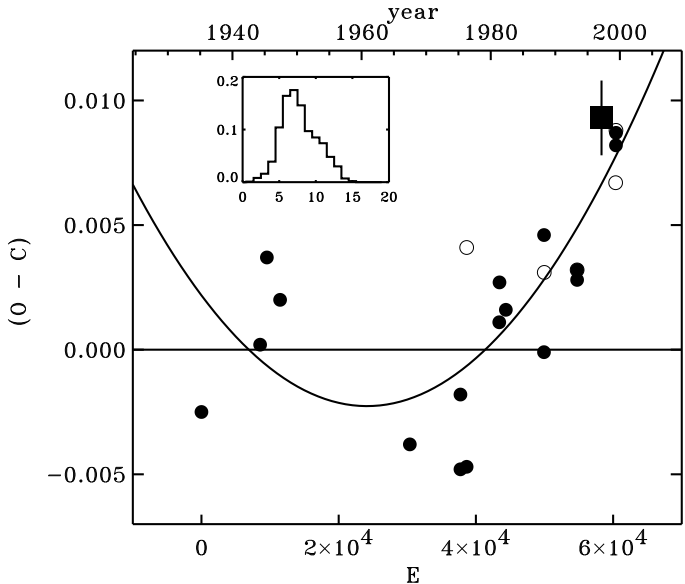,height=7cm}}
\vspace{0.25cm}
\caption{\label{fig6} The $(O-C)$ deviations in the
observed moments of eclipses (in days) from the ephemeris of Odell 
(1996). The secondary minima are marked by open
circles. The new spectroscopic determination is marked by
the large filled square. Its error has been obtained by
a bootstrap experiment and is well determined, but --
obviously -- systematic effects in photometric and
spectroscopic determinations may be different. The quadratic fit
discussed in the text is shown by a continuous line. The histogram
of the bootstrap results for the quadratic coefficient $a_2$
(in units of $10^{-12}$ days) is shown by the small insert. 
}
\end{figure}

Although known for almost 65 years, W Crv has not 
been extensively observed for moments of minima. Practically all 
extant data have been presented by Odell \shortcite{ode96}.
Dr.\ Odell kindly sent very new, unpublished data and corrections to
a few data points listed in Table~1 of his paper. These are given
in Table~\ref{tab4}. We have added to 
these the moment of minimum inferred from
our new spectroscopic determination of $T_0$ (see Table~\ref{tab2}).
I what follows, we will use the ephemeris of Odell:
$JD({\rm min}) = 2427861.3635 + 0.388080834 \times E$.
The {\it observed minus calculated\/} $(O-C)$ deviations
from Odell's ephemeris are shown in Figure~\ref{fig6}. The moments
secondary minima, which are based on shallower eclipses with
stronger light-curve perturbations, are marked in the figure
by open circles. Our spectroscopic result
gives a significant, positive deviation of 
$(O-C) = +0.0093 \pm 0.0015$ days, in agreement with the newest data
of Odell.
 
\begin{table}    
\caption{\label{tab4} New and corrected moments of 
minima for W~Crv}
\begin{tabular}{@{}cccc}
$E$     & $T_0$      & $(O-C)$  & Comment \\
        & 2400000+   &  days    &         \\
54750.0 & 49108.7920 & +0.0028  & correction \\
54752.5 & 49109.7626 & +0.0032  & correction \\
58309.0 & 50489.9781 & +0.0093  & spectroscopy \\
60364.5 & 51287.6757 & +0.0067  & new \\
60411.0 & 51305.7230 & +0.0082  & new \\
60413.5 & 51306.6938 & +0.0088  & new \\
60416.0 & 51307.6639 & +0.0087  & new \\
\end{tabular}
\end{table}

\begin{table}    
\caption{\label{tab5} Quadratic fits to the time-of-minima
$(O-C)$ deviations and the evolutionary time scales $\tau$} 
\begin{tabular}{@{}ccccc}
Value   &   $a_0$   &   $a_1$    &  $a_2$   &  $\tau$  \\
      &  days  &  $10^{-7}$ days & $10^{-12}$ days & $10^7$ years \\
$-95$\% ($-2\,\sigma$) & $-0.0025$ & $-8.28$ & $+3.87$ &   5.33 \\
$-68$\% ($-1\,\sigma$) & $-0.0002$ & $-6.44$ & $+5.93$ &   3.48 \\
median                & $+0.0023$ & $-3.79$ & $+7.98$ &   2.58 \\
$+68$\% ($+1\,\sigma$) & $+0.0072$ & $-2.36$ & $+11.06$ &  1.86 \\
$+95$\% ($+2\,\sigma$) & $+0.0097$ & $-1.04$ & $+13.44$ &  1.53 \\
\end{tabular}
\end{table}

The available times-of-minima contain information about
orbital period changes that have taken place over 
the 65 years. Disregarding presumably random and much smaller
shifts in the eclipse centres
caused by stellar-surface perturbations (whether we call them spots
or mass-transfer affected areas), the observed deviations from the linear
elements of Odell \shortcite{ode96} in Figure~\ref{fig6} 
can be interpretted as consisting of at least two streight segments
or as forming a parabola. We do not consider a possibility that
the discoverer of W~Crv, Tsesevich \shortcite{tse54},
committed a gross error in the timing of the minima
because he was one of the most experienced
observers of variable stars ever. In W~UMa-type systems, 
the abrupt changes of the type leading to the streight-segmented
$(O-C)$ diagrams take place in intervals of typically years; these changes
may have some relation to the magnetic-activity
cycles \cite{ruc85}. They are very difficult to handle as they require
very dense eclipse-timing coverage; such a coverage is not
available for W~Crv. It is easier to
analyze the $(O-C)$ deviations for a global quadratic trend 
using an expression: $(O-C) = a_0 + a_1 \times E + a_2 \times E^2$.
Because of poor distribution of data points over time, the linear
least-squares would give unreliable error estimates for the
coefficients $a_i$. In view of this difficulty,
the uncertainties have been evaluated using
the bootstrap-sampling technique and are listed
in Table~\ref{tab5} in terms of the median values at the
68 percent (for gaussian distributions, $\pm 1$-sigma) and  
95 percent ($\pm 2$-sigma) confidence levels. 
The bootstrap technique reveals a strongly non-gaussian
distribution of the uncertainties, as shown for the coefficient $a_2$
in the insert to Figure~\ref{fig6}.

The quadratic coefficient $a_2$ is proportional to the
second derivative of the times of minima hence to the period
change through $dP/dt = 2 a_2/P$. For comparison with the
theory of stellar evolution, it is convenient to consider the
{\it time-scale of the period change\/} given by
$\tau = P/(dP/dt) = P^2/2a_2$. The values of $\tau$ are
given in the last column of Table~\ref{tab5}. 
The data given in Table~\ref{tab5} indicate that the 
orbital period is becoming longer with the characteristic
time scale of $(1.5 - 5.3) \times 10^7$ years, with the range 
based on the highly secure 95 percent confidence level. The sense 
of the period change is somewhat unexpected as it indicates -- for the 
relative masses that we determined -- that the mass transfer
is from the less-massive component to the more-massive component,
i.e.\ as in Algols (the configuration designated as SC). 
One would normally expect
the other semi-detached configuration (SH) for
the pre-contact or broken-contact phases of the TRO cycles.
The period-lengthening argument for the Algol (SC)
configuration is a stronger one than any based on the light
curve analysis which seems to be hopelessly difficult for
W~Crv. The time-scale is exactly in the range expected for 
the Kelvin-Helmholtz or thermal time-scale evolution of solar-mass stars,
$\tau_{K-H} = 3.1 \times 10^7 (M/M_\odot)^2 
(R/R_\odot)^{-1} (L/L_\odot)^{-1}$,
which is characteristic for systems in the rapid stage of mass exchange
such as $\beta$~Lyrae or SV~Cen.

\section{Discussion and conclusions}
\label{disc}

The present paper contains results of spectroscopic observations
confirming the assumption of Odell \shortcite{ode96} that the
more massive, hotter star is eclipsed in the primary minimum. However,
this information and the value of the mass-ratio
are not sufficient to understand the exact nature
of the system mostly because of the strong light curve variability
which may be interpreted as an indication of mass-exchange and
accretion phenomena producing strong deviations from 
the standard binary-star model.
We suggest -- on the basis of the absence of light-curve
perturbations within the primary minima -- that the system is not a 
contact binary with components which mysteriously have 
different temperatures, but rather a semi-detached system. Furthermore,
we suggest that W~Crv, similarly to 
systems like V361~Lyr or SV~Cen, has a light-producing
volume between the stars or -- more likely -- on the inner
face of the secondary component. In the case of V361~Lyr,
there is apparently enough space for the stream of matter to be deflected 
by the Coriolis force and strike
the less-massive on the side; in SV~Cen, the photometric effects of
a strong contact are probably entirely due to the additional
light visible only in the orbital quadratures. In 
contrast to V361~Lyr and SV~Cen, the
mass-transfer phenomena in W~Crv are visible at all orbital phases 
except at primary minima, that is when the inner side of the
cooler component is directed away from the observer. 

The general considerations of the light-curve fits in the presence
of large brightness perturbations make both semi-detached 
configurations almost equally likely, 
but the semi-detached configuration of the Algol type for W~Crv, 
i.e.\ the one with the less-massive, 
cooler component filling the Roche lobe (SC) 
is preferable for two reasons: (1)~it is simpler, as it
leads to only one component deviating from the main-sequence relation
(since the inclination must be close to 90 degrees,
the secondary would have $0.92 R_\odot$ and $0.68 M_\odot$, whereas
the primary would be a solar-type star with
 $1.01 R_\odot$ and $1.00 M_\odot$), and
(2)~it can explain the observed {\it lengthening\/} of the orbital period
in the thermal time-scale. This way, W~Crv joins a group of well-known
stars -- such as SV~Cen, V361~Lyr or the famous $\beta$~Lyrae -- where 
large, systematic period changes are actually the final proof
of our hypothesis of the Algol configuration. W~Crv would be then the 
shortest-period (0.388 days) known Algol consisting of normal
(non-degenerate) components. With such a short period, the system presents 
a difficulty to the current theories 
describing formation of low-mass Algols, as reviewed by 
Yungelson et al.\ \shortcite{yun89}, and of binaries related to
contact systems, as reviewed by Eggleton \shortcite{egg96}.
One can only note that Sarna \& Fedorova \shortcite{SF89}, who considered
formation of solar-type contact binaries 
through the Case~A mass-exchange mechanism,
pointed out the importance of the initial mass-ratio: For mass-ratio 
sufficiently close to unity, the rapid (hydrodynamical) mass exchange
can be avoided and the system may evolve in 
the thermal time-scale. Although
the mass-reversal has not been modeled, 
it is likely that W~Crv is the product of such a process.

\section*{Acknowledgments}

We thank Dr.\ Andy Odell for providing the light curve and 
time-of-minima data and for extensive correspondence, numerous
advices and suggestions and Drs.\ Bohdan Paczy\'nski and
Janusz Ka\l u\.zny 
for a critical reading of the original version of the 
paper and several suggestions that improved the 
presentation of the paper.

\label{lastpage}
 
\end{document}